\documentclass[12 pt]{article} 
\usepackage{times,bm}
\usepackage[affil-it,]{authblk}
\usepackage[colorlinks,linkcolor=blue,citecolor=blue]{hyperref}
\usepackage[margin=0.5 in]{geometry}
\usepackage{graphicx,fancyref}
\usepackage{cite}
\usepackage{amsmath}
\usepackage{amsfonts}
\usepackage{amssymb}
\usepackage{color}
\usepackage{float}
\usepackage{multirow,multicol}
\usepackage{tabulary}
\usepackage[flushleft]{threeparttable}
\usepackage{pgfplots,subfigure}
\usepackage{xcolor}
\numberwithin{equation}{section}
\usepackage{tikz}
\usepgflibrary{arrows}
\usetikzlibrary{shapes.callouts}
\tikzset{  
	level/.style   = { thick, },
	connect/.style = { dotted, red   },
	notice/.style  = { draw, rectangle callout, callout relative pointer={#1} },
	label/.style   = { text width=2cm }
}

\begin{document}

\title{ \emph{q-Deformed Relativistic Fermion Scattering }}

\author[1]{\textit{Hadi Sobhani}%
	\thanks{Electronic address: hadisobhani8637@gmail.com (Corresponding author)}}
\affil[1]{{\small Physics Department, Shahrood University of Technology, Shahrood, Iran \\ P. O. Box: 3619995161-316.}}

\author[2]{\textit{Won Sang Chung}}
\affil[2]{{\small Department of Physics and Research Institute of Natural Science,
		College of Natural Science, Gyeongsang National University, Jinju 52828, Korea }}

\author[1]{\textit{Hassan Hassanabadi}}

\maketitle

\begin{abstract}
  \textit{In this article, after introducing a kind of q-deformation in quantum mechanics, first, q-deformed form of Dirac equation in relativistic quantum mechanics is derived. Then three important scat erring problem in physics are studied. All results have satisfied what we had expected before. Furthermore, effects of all parameters in the problems on the reflection and transmission coefficients are calculated and shown graphically.  }
\end{abstract}

\begin{small}
\end{small}

\begin{small}
\end{small}

\newpage
\section{Introduction}

q-deformation for quantum group and physical system has been one of remarkable and interesting issue of studies such as conformal quantum mechanics \cite{1} nuclear and high energy physics\cite{2,3,4} cosmic string and black holes \cite{5} fractional quantum Hall effect \cite{6}. Applications of q-deformation were emerged in physics and chemistry after introducing q-deformed harmonic oscillator \cite{7,8}  such as investigation of  electronic conductance in disordered metals and doped semiconductors \cite{9}, analyzing of the phonon spectrum in $^4$He \cite{10}, expressing of the oscillatory-rotational spectra of diatomic and multi-atomic molecules \cite{11,12}. Basically, q-calculus was established for the first time by Jackson \cite{13} then  Arik and Coon used it. 
Arik and Coon by studied generalized coherent states that are associated with a generalization of the harmonic oscillator commutation relation \cite{14}. They utilized
\begin{align}
	a{a^\dag } - q{a^\dag }a = 1,\quad \left[ {N,{a^\dag }} \right] = {a^\dag },\quad \left[ {N,a} \right] =  - a
\end{align}
where relation between  number operator and step operators ig given by
\begin{align}
	{a^\dag }a = {\left[ N \right]_q},
\end{align}
where a q-number is defined as 
\begin{align}
	{\left[ X \right]_q} = \frac{{1 - {q^X}}}{{1 - q}}.
\end{align}

Another q-deformation exists that has been introduced by Tsallis \cite{Ts} and has a different algebraic structure from Jackson's. For the Tsallis's case the q-derivative and q-integral was given by Borges \cite{15}. Recently Chung and Hassanabadi \cite{16}  used the q-derivative emerging in the non-extensive statistical physics to formulate the q-deformed quantum mechanics. 

In what follows: Sec. 2 is devoted to introduction to the kind of q-deformation of quantum mechanics which will be used in the next sections. In Sec. 3, q-deformed version of Dirac equation is derived. As first relativistic scattering problem in q-deformed version of relativistic quantum mechanics, scattering from a Dirac delta potential is done in Sec. 4. Sec. 5 is devoted to extended form of problem in Sec. 4, scattering problem from a double Dirac delta potential. In the last Ramsauer-Townsend effect is studied in considered formalism of quantum mechanics.

\section{q-Deformed Quantum Mechanics}

In this section, we want to introduce postulates of q-deformed quantum mechanics to use for the next sections. In this formalism of quantum mechanics we deal with:  
\begin{enumerate}
	\item 
	In this formalism of quantum mechanics like the ordinary one, time-dependent form of Schr\"{o}dinger equation in q-deformed quantum mechanics is written in form of
	\begin{align}
		\label{2-1}
		i\hbar \frac{{\partial \psi \left( {x,t} \right)}}{{\partial t}} = H\left( {\hat x,\hat p} \right)\psi \left( {x,t} \right) = \left( {\frac{{{p^2}}}{{2m}} + V\left( {\hat x} \right)} \right)\psi \left( {x,t} \right),
	\end{align}
	in which we deal with the operators as
	\begin{align}
		\label{2-2}
		\hat p =  - i\hbar {D_x} =  - i\hbar \left( {1 + qx^2} \right)\frac{d}{{dx}} \qquad \hat x = x,
	\end{align}
	where $q$ is a positive constant and the wave function  $\psi \left( {x,t} \right)$.
	
	\item
	Inner product of Hilbert space in one-dimensional q-deformed quantum mechanics can be written as
	\begin{equation}
		\label{2-3}
		\begin{gathered}
			\left\langle {f}
			\mathrel{\left | {\vphantom {f g}}
				\right. \kern-\nulldelimiterspace}
			{g} \right\rangle  = \int\limits_{ - \infty }^\infty  {{g^*}\left( x \right)f\left( x \right){d_q}x} , \hfill \\
			{d_q}x = \frac{{dx}}{{\left( {1 + qx^2} \right)}}. \hfill \\ 
		\end{gathered} 
	\end{equation}	
	
	\item
	Expectation value of an operator $\hat{O}$ regarding the wave function $\psi \left( {x,t} \right)$ is given by 
	\begin{align}
		\label{2-4}
		\left\langle \hat{O} \right\rangle  = \left\langle {\psi }
		\mathrel{\left | {\vphantom {\psi  {\hat{O} \psi }}}
			\right. \kern-\nulldelimiterspace}
		{{\hat{O}\psi }} \right\rangle  = \int\limits_{-\infty}^{\infty} {{\psi ^*}\left( {x,t} \right)\hat{O}\psi \left( {x,t} \right)} {d_q}x,
	\end{align}
	also we have Hermitian definition for the operator if we get
	\begin{align}
		\label{2-5}
		\left\langle {\psi }
		\mathrel{\left | {\vphantom {\psi  {\hat{O}\psi }}}
			\right. \kern-\nulldelimiterspace}
		{{\hat{O}\psi }} \right\rangle  = \left\langle {{\hat{O}\psi }}
		\mathrel{\left | {\vphantom {{\hat{O}\psi } \psi }}
			\right. \kern-\nulldelimiterspace}
		{\psi } \right\rangle .
	\end{align}
\end{enumerate}
Remarking this point is suitable that we have been considering deformation only for coordinate part, then the time part has no deformation. This point can be checked in the first postulate.

In this formalism of quantum mechanics, commutation relation between coordinate and its momentum should be deformed in form of
\begin{align}
	\label{2-6}
	\left[ {\hat x,\hat p} \right] = i\hbar \left( {1 + q \hat x ^2} \right).
\end{align}
Considering operator form of coordinate and momentum
\begin{align}
	\label{2-7}
	\hat x \leftrightarrow x,\qquad \hat p \leftrightarrow  - i\hbar {D_x},
\end{align} 
we can rewrite Eq.\eqref{2-1} in terms of the operators
\begin{align}
	\label{2-8}
	i\hbar \frac{{\partial \psi \left( {x,t} \right)}}{{\partial t}} = \left( {\frac{{ - {\hbar ^2}}}{{2m}}D_x^2 + V\left( x \right)} \right)\psi \left( {x,t} \right),
\end{align}
and to obtain time-independent form of Shcr\"{o}dinger equation in this formalism, we set $\psi \left( {x,t} \right) = {e^{\frac{{ - i}}{\hbar }Et}}\phi \left( x \right)$ then we have 
\begin{align}
	\label{2-9}
	\left( {\frac{{ - {\hbar ^2}}}{{2m}}D_x^2 + V\left( x \right)} \right)\phi \left( x \right) = E\phi \left( x \right).
\end{align}

Using \eqref{2-8}, we can easily find continuity relation in this formalism of quantum mechanics as 
\begin{align}
	\label{2-10}
	\frac{{\partial \rho \left( {x,t} \right)}}{{\partial t}} + {D_x}j\left( {x,t} \right) = 0,
\end{align}
where
\begin{align}
	\label{2-11}
	&\rho \left( {x,t} \right) = {\psi ^*}\left( {x,t} \right)\psi \left( {x,t} \right), \hfill \\
	\label{2-12}
	&j\left( {x,t} \right) = \frac{\hbar }{{2mi}}\left( {{\psi ^*}\left( {x,t} \right){D_x}\psi \left( {x,t} \right) - \psi \left( {x,t} \right){D_x}{\psi ^*}\left( {x,t} \right)} \right). \hfill 
\end{align} 

By these considerations, we are in a position to study relativistic scattering of fermions in q-deformed relativistic quantum mechanics. 

\section{Scattering of Relativistic Fermions in q-Deformed Quantum Mechanics }

In this section, we want to study scattering of fermions in q-deformed formalism of relativistic quantum mechanics. Study of fermions can be done by  Dirac equation. This can be written as $\hbar=c=1$ \cite{17}
\begin{align}
\label{3-1}
i\frac{{\partial \Psi \left( {x,t} \right)}}{{\partial t}} = \left( {{\boldsymbol{\alpha }}.{\bf{p}} + \beta \left( {m + S\left( x \right)} \right) + V\left( x \right)} \right)\Psi \left( {x,t} \right),
\end{align}
in which the matrices are
\begin{align}
\label{3-2}
{\boldsymbol{\alpha }} = \left( {\begin{array}{*{20}{c}}
	0&{\boldsymbol{\sigma }}\\
	{\boldsymbol{\sigma }}&0
	\end{array}} \right),\quad \beta  = \left( {\begin{array}{*{20}{c}}
	1&0\\
	0&{ - 1}
	\end{array}} \right)
\end{align}
where $\sigma$ stands for Pauli matrices. We have considered $x$ direction as interaction direction. to obtain stationary states, we choose the wave function as 
\begin{align}
\label{3-3}
\Psi \left( {x,t} \right) = {e^{ - iEt}}\Phi \left( x \right) = {e^{ - iEt}}\left( {\begin{array}{*{20}{c}}
	{{\Phi _u}\left( x \right)}\\
	{{\Phi _d}\left( x \right)}
	\end{array}} \right),
\end{align} 
also, we'd like to consider $S\left( x \right) = V\left( x \right)$ for simplicity. These assumptions give us a system of equation like
\begin{align}
\label{3-4}
&\left( {m + 2V\left( x \right) - E} \right){\Phi _u}\left( x \right) + {\sigma _x}\left( { - i\left( {1 + q{x^2}} \right)\frac{{d{\Phi _d}\left( x \right)}}{{dx}}} \right) = 0,\\
\label{3-5}
&{\sigma _x}\left( { - i\left( {1 + q{x^2}} \right)\frac{{d{\Phi _u}\left( x \right)}}{{dx}}} \right) - \left( {E + m} \right){\Phi _d}\left( x \right) = 0.
\end{align}
From Eq. \eqref{3-5} we find that
\begin{align}
\label{3-6}
{\Phi _d}\left( x \right) = {\sigma _x}\frac{{ - i\left( {1 + q{x^2}} \right)}}{{E + m}}\frac{{d{\Phi _u}\left( x \right)}}{{dx}},
\end{align}
If one substitutes Eq. \eqref{3-6} into Eq. \eqref{3-4}, easily can derive
\begin{align}
\label{3-7}
&{\left( {1 + q{x^2}} \right)^2}\frac{{{d^2}{\Phi _u}\left( x \right)}}{{d{x^2}}} + 2qx\left( {1 + q{x^2}} \right)\frac{{d{\Phi _u}\left( x \right)}}{{dx}} + \left( {{p^2} - 2\left( {E + m} \right)V\left( x \right)} \right){\Phi _u}\left( x \right) = 0,\\
\label{3-8}
&{p^2} = {E^2} - {m^2}
\end{align}

In the next sections, we will study three important and famous scattering.

\section{Scattering Due to Single Dirac Delta Potential }

As first scattering study, we want to consider single Dirac delta potential as
\begin{align}
\label{4-1}
V\left( x \right) = {V_1}\delta \left( {x - a_1} \right)
\end{align}
where $V_1$ and $a_1$ are real constants. This point can be derived that this potential produces a discontinuity for first derivative of wave function as
\begin{align}
\label{4-2}
\frac{{d{\Phi _u}\left( {x = {a_1 ^ + }} \right)}}{{dx}} - \frac{{d{\Phi _u}\left( {x = {a_1 ^ - }} \right)}}{{dx}} = \frac{{2\left( {m + E} \right)}}{{{{\left( {1 + q{a_1 ^2}} \right)}^2}}}{\Phi _u}\left( {x = a_1} \right)
\end{align}  
We assume that particles come from $x<a_1$ then because of Dirac delta existence they scatter. Consequently some them reflected to region I $\left( x<a_1\right) $ and the other transmit to region II $\left( x>a_1\right)$. According to this assumption, we can find wave functions of the regions as
\begin{align}
\label{4-3}
&{\Phi _{u,I}}\left( x \right) = {e^{ip\left( {\frac{1}{{\sqrt q }}{\mathop{\rm tan^{-1}}\nolimits} \left( {\sqrt q x} \right)} \right)}} + r{e^{ - ip\left( {\frac{1}{{\sqrt q }}{\mathop{\rm tan^{-1}}\nolimits} \left( {\sqrt q x} \right)} \right)}},\\
\label{4-4}
&{\Phi _{u,II}}\left( x \right) = t{e^{ip\left( {\frac{1}{{\sqrt q }}{\mathop{\rm tan^{-1}}\nolimits} \left( {\sqrt q x} \right)} \right)}}.
\end{align}
The coefficients $r_1$ and $t_1$ can be determined by using boundary condition of continuity and discontinuity of wave functions at $x=a_1$. These are
\begin{align}
\label{4-5}
&{e^{ip\left( {\frac{1}{{\sqrt q }}{\mathop{\rm tan^{-1}}\nolimits} \left( {\sqrt q a_1} \right)} \right)}} + r{e^{ - ip\left( {\frac{1}{{\sqrt q }}{\mathop{\rm tan^{-1}}\nolimits} \left( {\sqrt q a_1} \right)} \right)}} = t{e^{ip\left( {\frac{1}{{\sqrt q }}{\mathop{\rm tan^{-1}}\nolimits} \left( {\sqrt q a_1} \right)} \right)}},\\
\label{4-6}
&{e^{ip\left( {\frac{1}{{\sqrt q }}{\mathop{\rm tan^{-1}}\nolimits} \left( {\sqrt q a_1} \right)} \right)}}\left( {t - 1} \right) + r{e^{ - ip\left( {\frac{1}{{\sqrt q }}{\mathop{\rm tan^{-1}}\nolimits} \left( {\sqrt q a_1} \right)} \right)}} = \frac{{2\left( {m + E} \right){V_1}}}{{\left( {1 + q{a_1 ^2}} \right)ip}}t{e^{ip\left( {\frac{1}{{\sqrt q }}{\mathop{\rm tan^{-1}}\nolimits} \left( {\sqrt q a_1} \right)} \right)}}.
\end{align}
where by solving them, we can find that
\begin{align}
&r_1= -\frac{V (e+m) e^{\frac{2 i p \tan ^{-1}\left(a_1 \sqrt{q}\right)}{\sqrt{q}}}}{V (e+m)-i p \left(a_1 ^2 q+1\right)}\\
&t_1 = \frac{a_1 ^2 p q+p}{a_1 ^2 p q+i V (e+m)+p}.
\end{align}	

\begin{figure}[H]
	\centering
	\includegraphics[scale=0.8]{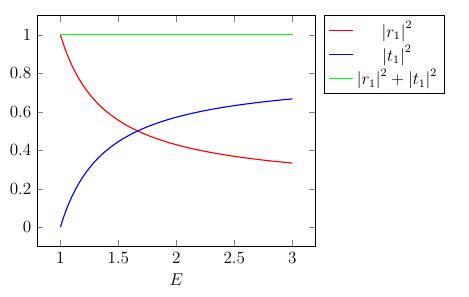}
	\caption{Plots of ${\left| r_1 \right|^2}$, ${\left| t_1 \right|^2}$ and ${\left| r_1 \right|^2}+{\left| t_1 \right|^2}$ as energy varies.  }
		\label{fig:1}
\end{figure}

On the other hand, current density of fermions can be derived by $j = {\Psi ^\dag }\alpha \Psi$. Because there is no sink or source, we have from current density
\begin{align}
\label{4-7}
{\left| r_1 \right|^2} + {\left| t_1 \right|^2} = 1.
\end{align}
We plot this equation considering $V_1=2,q=3,a_1=1$ and $m=1$ in figure \ref{fig:1}.

\begin{figure}[H]
	\centering
	\includegraphics{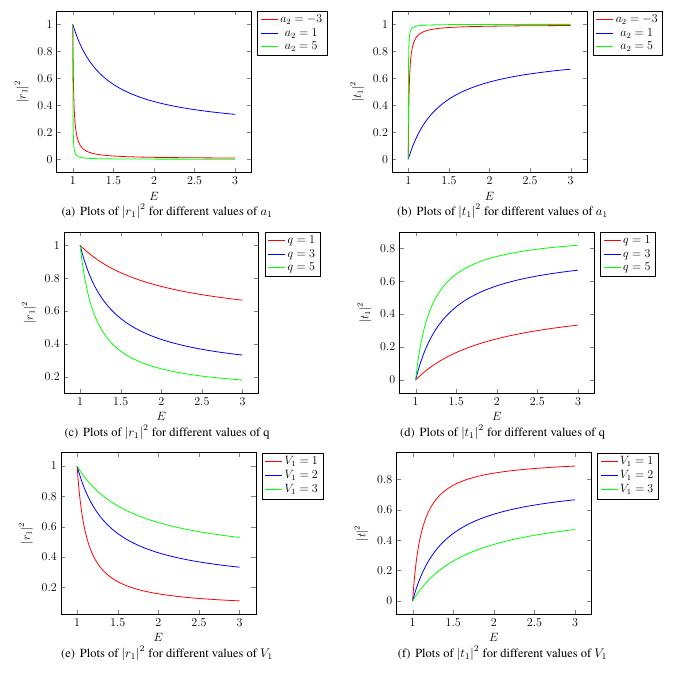}
	\caption{In this figure, different treatments of ${\left| t_1 \right|^2}$ and ${\left| r_1 \right|^2}$ as parameters $a_1, q$ and $V_1$ vary, have been plotted. We set the parameters in (a) and (b) $V_1=2$, $q=3$, (c) and (d) $V_1=2$, $a_1=1$, (e) and (f) $q=3$, $a_1=1$. }
	\label{fig:2}
\end{figure}

For further study about effects of different parameters on the reflection and transmission coefficients, we have plot figure \ref{fig:2} in which readers can see that how different parameters in our system can affect of these confinements

As it can be seen that there no fluctuations in the figures and each curve has a smooth treatment. But in the next section we will deal with interesting. Because kind of scatter potential makes interesting results resemble one of most important and famous effects in physics.

\section{Scattering From Double Dirac Delta Potential}

In this section, we suppose that particle are scattered from a double Dirac delta potential in form of
\begin{align}
\label{5-1}
V\left( x \right) = {V_2}\left( {\delta \left( {x + a_2} \right) + \delta \left( {x - a_2} \right)} \right),
\end{align}
where $V_2$ and $a_2$ are real constants. From the previous section we know that this kind of potential makes discontinuity for derivative of wave functions at $x=a_2$ and $x=-a_2$. They are
\begin{align}
\label{5-2}
&\frac{{d{\Phi _u}\left( {x = {a_2 ^ + }} \right)}}{{dx}} - \frac{{d{\Phi _u}\left( {x = {a_2 ^ - }} \right)}}{{dx}} = \frac{{2\left( {m + E} \right)}}{{{{\left( {1 + q{a_2 ^2}} \right)}^2}}}{\Phi _u}\left( {x = a_2} \right),x = a_2,\\
\label{5-3}
&\frac{{d{\Phi _u}\left( {x =  - {a_2 ^ + }} \right)}}{{dx}} - \frac{{d{\Phi _u}\left( {x =  - {a_2 ^ - }} \right)}}{{dx}} = \frac{{2\left( {m + E} \right)}}{{{{\left( {1 + q{a_2 ^2}} \right)}^2}}}{\Phi _u}\left( {x =  - a_2} \right),x =  - a_2.
\end{align}
Similar the previous section, we suppose that particles come from region I $\left(  x<-a_2 \right) $. Then they are scattered in region II $\left( -a_2<x<a_2 \right) $. So some of them will be reflected into region I and the other will be transmitted into region III $\left(  x> a_2 \right) $. According to this assumption, we have wave functions
\begin{align}
\label{5-4}
&{\Phi _{u,I}}\left( x \right) = {e^{ip\left( {\frac{1}{{\sqrt q }}{\mathop{\rm tan^{-1}}\nolimits} \left( {\sqrt q x} \right)} \right)}} + r{e^{ - ip\left( {\frac{1}{{\sqrt q }}{\mathop{\rm tan^{-1}}\nolimits} \left( {\sqrt q x} \right)} \right)}},\\
\label{5-5}
&{\Phi _{u,II}}\left( x \right) = A{e^{ip\left( {\frac{1}{{\sqrt q }}{\mathop{\rm tan^{-1}}\nolimits} \left( {\sqrt q x} \right)} \right)}} + B{e^{ - ip\left( {\frac{1}{{\sqrt q }}{\mathop{\rm tan^{-1}}\nolimits} \left( {\sqrt q x} \right)} \right)}},\\
\label{5-6}
&{\Phi _{u,III}}\left( x \right) = t{e^{ip\left( {\frac{1}{{\sqrt q }}{\mathop{\rm tan^{-1}}\nolimits} \left( {\sqrt q x} \right)} \right)}}.
\end{align}
where the coefficients are constant which can be determined from continuity and discontinuity condition at $x=a_2$ and $x=-a_2$. Using these conditions lead to the system of four equations

\begin{equation}
\label{5-7}
\left\{ \begin{array}{l}
\left. \begin{array}{l}
{e^{ip\left( {\frac{1}{{\sqrt q }}{\mathop{\rm tan^{-1}}\nolimits} \left( { - \sqrt q a_2} \right)} \right)}} + r{e^{ - ip\left( {\frac{1}{{\sqrt q }}{\mathop{\rm tan^{-1}}\nolimits} \left( { - \sqrt q a_2} \right)} \right)}} = A{e^{ip\left( {\frac{1}{{\sqrt q }}{\mathop{\rm tan^{-1}}\nolimits} \left( { - \sqrt q a_2} \right)} \right)}} + B{e^{ - ip\left( {\frac{1}{{\sqrt q }}{\mathop{\rm tan^{-1}}\nolimits} \left( { - \sqrt q a_2} \right)} \right)}},\\
{e^{ip\left( {\frac{1}{{\sqrt q }}{\mathop{\rm tan^{-1}}\nolimits} \left( { - \sqrt q a_2} \right)} \right)}}\left( {A - 1} \right) + {e^{ - ip\left( {\frac{1}{{\sqrt q }}{\mathop{\rm tan^{-1}}\nolimits} \left( { - \sqrt q a_2} \right)} \right)}}\left( {r - B} \right)\\
= \frac{{2{V_2}\left( {E + m} \right)}}{{ip\left( {1 + q{a_2 ^2}} \right)}}\left( {{e^{ip\left( {\frac{1}{{\sqrt q }}{\mathop{\rm tan^{-1}}\nolimits} \left( { - \sqrt q a_2} \right)} \right)}} + r{e^{ - ip\left( {\frac{1}{{\sqrt q }}{\mathop{\rm tan^{-1}}\nolimits} \left( { - \sqrt q a_2} \right)} \right)}}} \right),
\end{array} \right\}x =  - a_2,\\
\left. \begin{array}{l}
A{e^{ip\left( {\frac{1}{{\sqrt q }}{\mathop{\rm tan^{-1}}\nolimits} \left( {\sqrt q a_2} \right)} \right)}} + B{e^{ - ip\left( {\frac{1}{{\sqrt q }}{\mathop{\rm tan^{-1}}\nolimits} \left( {\sqrt q a_2} \right)} \right)}} = t{e^{ip\left( {\frac{1}{{\sqrt q }}{\mathop{\rm tan^{-1}}\nolimits} \left( {\sqrt q a_2} \right)} \right)}},\\
{e^{ip\left( {\frac{1}{{\sqrt q }}{\mathop{\rm tan^{-1}}\nolimits} \left( {\sqrt q a_2} \right)} \right)}}\left( {t - A} \right) + B{e^{ - ip\left( {\frac{1}{{\sqrt q }}{\mathop{\rm tan^{-1}}\nolimits} \left( {\sqrt q a_2} \right)} \right)}}\\
= \frac{{2{V_2}\left( {E + m} \right)}}{{ip\left( {1 + q{a_2 ^2}} \right)}}\left( {t{e^{ip\left( {\frac{1}{{\sqrt q }}{\mathop{\rm tan^{-1}}\nolimits} \left( {\sqrt q a_2} \right)} \right)}}} \right),
\end{array} \right\}x = a_2.
\end{array} \right.
\end{equation}
Solving this system of equations for the constants of wave functions we obtain
\begin{align}
&r_2=-\frac{V (e+m) e^{-\frac{2 i p \tan ^{-1}\left(a_2 \sqrt{q}\right)}{\sqrt{q}}} \left(e^{\frac{4 i p \tan ^{-1}\left(a_2 \sqrt{q}\right)}{\sqrt{q}}} \left(V
	(e+m)+i p \left(a_2 ^2 q+1\right)\right)+i \left(a_2 ^2 p q+i m V+p\right)-e V\right)}{\left(a_2 ^2 p q+i V (e+m)+p\right)^2+V^2 (e+m)^2 e^{\frac{4 i p \tan
			^{-1}\left(a_2 \sqrt{q}\right)}{\sqrt{q}}}}\\
&t_2=\frac{\left(a_2 ^2 p q+p\right)^2}{\left(a_2 ^2 p q+i V (e+m)+p\right)^2+V^2 (e+m)^2 e^{\frac{4 i p \tan ^{-1}\left(a_2 \sqrt{q}\right)}{\sqrt{q}}}}
\end{align}

\begin{figure}[H]
	\centering
	\includegraphics{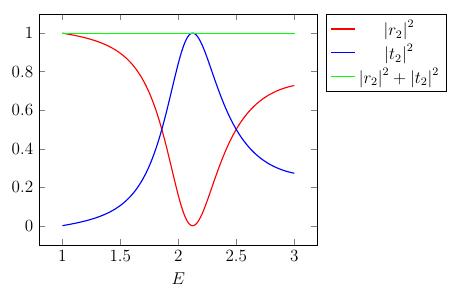}
	\caption{Plots of ${\left| r_2 \right|^2}$, ${\left| t_2 \right|^2}$ and ${\left| r_2 \right|^2}+{\left| t_2 \right|^2}$ as energy varies.}
	\label{fig:3}
\end{figure}

Using the similar manner of previous section, we have found the constraint 
\begin{align}
\label{5-8}
{\left| {{r_2}} \right|^2} + {\left| {{t_2}} \right|^2} = 1.
\end{align}
By plotting this equation using Eq. \eqref{5-7} we can check validity of it. This point can be seen in figure \ref{fig:3}

\begin{figure}[H]
	\centering
	\includegraphics[scale=0.9]{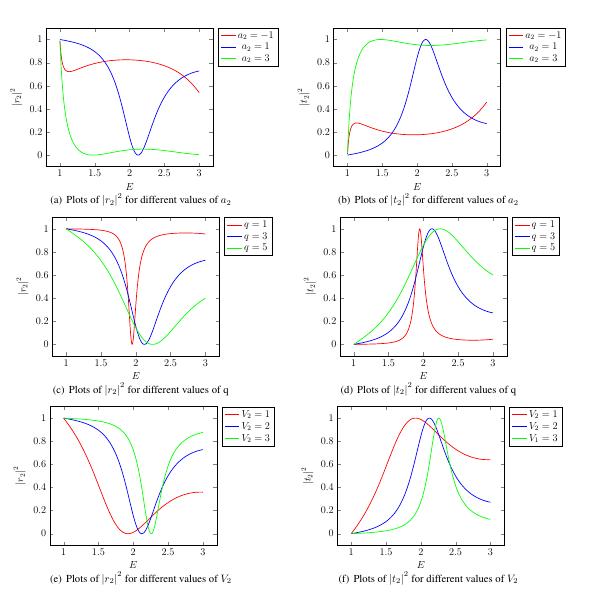}
	\caption{In this figure, different treatments of ${\left| t_2 \right|^2}$ and ${\left| r_2 \right|^2}$ as parameters $a_2, q$ and $V_2$ vary, have been plotted. We set the parameters in (a) and (b) $V_2=2$, $q=3$, (c) and (d) $V_2=2$, $a_2=1$, (e) and (f) $q=3$, $a_2=1$.  }
		\label{fig:4}
\end{figure}

It is instructive if we check treatments of reflection and transmission coefficients as different parameters in vary. This one is done and plotted in figure \ref{fig:4}. 

The main difference between this section and the previous section is shown in the results. In this section by considering a double Dirac potential as scatter potential, we find out some fluctuation in reflection and transmission coefficients, however in the previous section we dealt with a smooth treatment. On the other hand, such fluctuations remember us one of most famous  and important effect in physics, Ramsauer-Townsend effect. In the next section we will investigate this effect in q-deformed relativistic version.

\section{Ramsauer-Townsend Effect in q-Deformed Relativistic Quantum Mechanics}

\begin{figure}[H]
	\centering
	\includegraphics{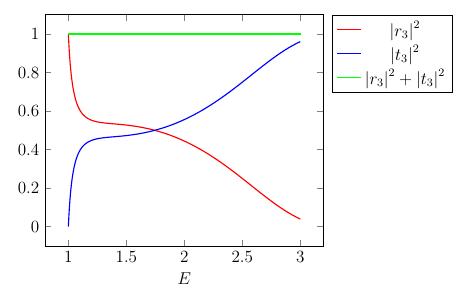}
	\caption{Plots of ${\left| r_3 \right|^2}$, ${\left| t_3 \right|^2}$ and ${\left| r_3 \right|^2}+{\left| t_3 \right|^2}$ as energy varies.}
	\label{fig:5}
\end{figure}

\begin{figure}[H]
	\centering
	\includegraphics[scale=0.8]{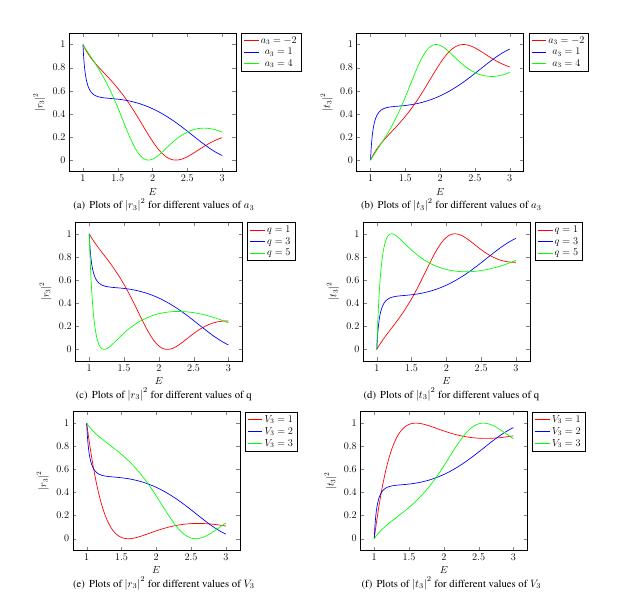}
	\caption{In this figure, different treatments of ${\left| t_3 \right|^2}$ and ${\left| r_3 \right|^2}$ as parameters $a_3, q$ and $V_3$ vary, have been plotted. We set the parameters in (a) and (b) $V_3=2$, $q=3$, (c) and (d) $V_3=2$, $a_3=1$, (e) and (f) $q=3$, $a_3=1$.}
	\label{fig:6}
\end{figure}

Ramsauer-Townsend effect is a face of electron scattering. This scattering due to a simple potential well. Actually when as electron which is moving through noble gas such as Xenon with low energy (like 0.1 $eV$) something strange happens. There an anomalously large transmission in this scattering\cite{1}. Importance of this simple scattering is that this effect can only be decried by quantum mechanics. 
Now, in considered formalism of relativistic quantum mechanics, we want to study. Considering a potential well such
\begin{align}
\label{6-1}
\begin{cases}
-V	\quad &for \quad -a_3<x<a_3,\\
0	\quad &for \quad elsewhere.
\end{cases}
\end{align}
and following previous assumptions, we can derived wave functions of our three regions in the problem as following
\begin{align}
\label{6-2}
&{\Phi _{u,I}}\left( x \right) = {e^{ip\left( {\frac{1}{{\sqrt q }}{\mathop{\rm tan^{-1}}\nolimits} \left( {\sqrt q x} \right)} \right)}} + {r_3}{e^{ - ip\left( {\frac{1}{{\sqrt q }}{\mathop{\rm tan^{-1}}\nolimits} \left( {\sqrt q x} \right)} \right)}},\\
\label{6-3}
&{\Phi _{u,II}}\left( x \right) = {A_3}{e^{i\eta \left( {\frac{1}{{\sqrt q }}{\mathop{\rm tan^{-1}}\nolimits} \left( {\sqrt q x} \right)} \right)}} + {B_3}{e^{ - i\eta \left( {\frac{1}{{\sqrt q }}{\mathop{\rm tan^{-1}}\nolimits} \left( {\sqrt q x} \right)} \right)}},\\
\label{6-4}
&{\Phi _{u,III}}\left( x \right) = {t_3}{e^{ip\left( {\frac{1}{{\sqrt q }}{\mathop{\rm tan^{-1}}\nolimits} \left( {\sqrt q x} \right)} \right)}},\\
\label{6-5}
&{\eta ^2} = {p^2} + 2\left( {E + m} \right)V.
\end{align}
in which the coefficients can be given by using continuity conditions of wave functions and their derivatives at $x=-a_3$ and $x=a_3$. Using these conditions results in

\begin{equation}
\label{6-6}
\left\{ \begin{array}{l}
\left. \begin{array}{l}
{e^{ip\left( {\frac{1}{{\sqrt q }}{\mathop{\rm tan^{-1}}\nolimits} \left( { - \sqrt q a_3} \right)} \right)}} + {r_3}{e^{ - ip\left( {\frac{1}{{\sqrt q }}{\mathop{\rm tan^{-1}}\nolimits} \left( { - \sqrt q a_3} \right)} \right)}} = {A_3}{e^{i\eta \left( {\frac{1}{{\sqrt q }}{\mathop{\rm tan^{-1}}\nolimits} \left( { - \sqrt q a_3} \right)} \right)}} + {B_3}{e^{ - i\eta \left( {\frac{1}{{\sqrt q }}{\mathop{\rm tan^{-1}}\nolimits} \left( { - \sqrt q a_3} \right)} \right)}},\\
p\left( {{e^{ip\left( {\frac{1}{{\sqrt q }}{\mathop{\rm tan^{-1}}\nolimits} \left( { - \sqrt q a_3} \right)} \right)}} - {r_3}{e^{ - ip\left( {\frac{1}{{\sqrt q }}{\mathop{\rm tan^{-1}}\nolimits} \left( { - \sqrt q a_3} \right)} \right)}}} \right)\\
= \eta \left( {{A_3}{e^{i\eta \left( {\frac{1}{{\sqrt q }}{\mathop{\rm tan^{-1}}\nolimits} \left( { - \sqrt q a_3} \right)} \right)}} - {B_3}{e^{ - i\eta \left( {\frac{1}{{\sqrt q }}{\mathop{\rm tan^{-1}}\nolimits} \left( { - \sqrt q a_3} \right)} \right)}}} \right),
\end{array} \right\}x = {a_3},\\
\left. \begin{array}{l}
{A_3}{e^{i\eta \left( {\frac{1}{{\sqrt q }}{\mathop{\rm tan^{-1}}\nolimits} \left( {\sqrt q a_3} \right)} \right)}} + {B_3}{e^{ - i\eta \left( {\frac{1}{{\sqrt q }}{\mathop{\rm tan^{-1}}\nolimits} \left( {\sqrt q a_3} \right)} \right)}} = {t_3}{e^{ip\left( {\frac{1}{{\sqrt q }}{\mathop{\rm tan^{-1}}\nolimits} \left( {\sqrt q a_3} \right)} \right)}},\\
\eta \left( {{A_3}{e^{i\eta \left( {\frac{1}{{\sqrt q }}{\mathop{\rm tan^{-1}}\nolimits} \left( {\sqrt q a} \right)} \right)}} - {B_3}{e^{ - i\eta \left( {\frac{1}{{\sqrt q }}{\mathop{\rm tan^{-1}}\nolimits} \left( {\sqrt q a_3} \right)} \right)}}} \right) = p{t_3}{e^{ip\left( {\frac{1}{{\sqrt q }}{\mathop{\rm tan^{-1}}\nolimits} \left( {\sqrt q a_3} \right)} \right)}},
\end{array} \right\}x = {a_3}.
\end{array} \right.
\end{equation}
From this system of equation we can determine the coefficients as
\begin{align}
&r_3 = \frac{(p-\eta ) (\eta +p) e^{-\frac{2 i p \tan ^{-1}\left(a_3 \sqrt{q}\right)}{\sqrt{q}}} \left(-1+e^{\frac{4 i \eta  \tan ^{-1}\left(a_3
\sqrt{q}\right)}{\sqrt{q}}}\right)}{p^2 e^{\frac{4 i \eta  \tan ^{-1}\left(a_3 \sqrt{q}\right)}{\sqrt{q}}}-2 \eta  p e^{\frac{4 i \eta  \tan ^{-1}\left(a_3
\sqrt{q}\right)}{\sqrt{q}}}+\eta ^2 e^{\frac{4 i \eta  \tan ^{-1}\left(a_3 \sqrt{q}\right)}{\sqrt{q}}}-\eta ^2-p^2-2 \eta  p},\\
&t_3=-\frac{4 \eta  p  e^{\left(\frac{2 i \eta  \tan ^{-1}\left(a_3 \sqrt{q}\right)}{\sqrt{q}}-\frac{2 i p \tan ^{-1}\left(a_3
\sqrt{q}\right)}{\sqrt{q}}\right)}}{p^2 e^{\frac{4 i \eta  \tan ^{-1}\left(a_3 \sqrt{q}\right)}{\sqrt{q}}}-2 \eta  p e^{\frac{4 i \eta  \tan ^{-1}\left(a_3
\sqrt{q}\right)}{\sqrt{q}}}+\eta ^2 e^{\frac{4 i \eta  \tan ^{-1}\left(a_3 \sqrt{q}\right)}{\sqrt{q}}}-\eta ^2-p^2-2 \eta  p}
\end{align}
 Like previous sections, by using definition of current density we can find out that the constraint
\begin{align}
\label{6-7}
{\left| {{r_2}} \right|^2} + {\left| {{t_2}} \right|^2} = 1,
\end{align} 
is governed here. By Solving Eq. \eqref{6-6} and  plotting Eq.\eqref{6-7} in figure \ref{fig:5} we have
which is similar to what we faced in the previous sections. Furthermore treatments of reflection and transmission coefficients in terms of different parameters  are plotted in  figure \ref{fig:6}.

\section{Conclusions}

In this article, we introduced a q-deformation of quantum mechanics. Then in such formalism of quantum mechanics, we studied three important and famous scattering problems in relativistic region. We first rewrote Dirac equation in q-deformation then as first case, scattering due to a single Dirac delta potential was studied. In this case we dealt with smooth treatments if reflection and transmission coefficients. In the next case, a double Dirac delta potential was considered. In this case we saw that there were some fluctuation in reflection and transmission coefficients which were similar to Ramsauer-Townsend effect. To check this point, we also investigated this effect in relativistic region. By plotting the coefficients we found out that effect of scattering from a potential well in q-deformed version of relativistic quantum mechanics could be simulated by considering double Dirac delta potential.

\textbf{Acknowledgment} \\
It is a great pleasure for the authors to thank the referee because of the helpful comments.


\end{document}